\documentclass[aps,prl,twocolumn,superscriptaddress]{revtex4-1}

\usepackage{graphics}
\usepackage{graphicx}
\usepackage{float}
\usepackage{amsmath}
\usepackage{amssymb}
\usepackage{color}

\begin{document}

\title{Direct measurement of the critical pore size in a model membrane }

\author{Mark Ilton}
\affiliation{Department of Physics \& Astronomy, McMaster University, Hamilton, Ontario, Canada, L8S 4M1}
\author{Christian DiMaria}
\affiliation{Department of Physics \& Astronomy, McMaster University, Hamilton, Ontario, Canada, L8S 4M1}
\author{Kari Dalnoki-Veress}
\email{dalnoki@mcmaster.ca}
\affiliation{Department of Physics \& Astronomy, McMaster University, Hamilton, Ontario, Canada, L8S 4M1}
\affiliation{Laboratoire de Physico-Chimie Th\'{e}orique, UMR CNRS Gulliver 7083, ESPCI ParisTech, PSL Research University, 75005 Paris, France}

\date{\today}

\begin{abstract}
We study pore nucleation in a model membrane system, a freestanding polymer film. Nucleated pores smaller than a critical size close, while pores larger than the critical size grow. Holes of varying size were purposefully prepared in liquid polymer films, and their evolution in time was monitored using optical and atomic force microscopy to extract a critical radius. The critical radius scales linearly with film thickness for a homopolymer film. The results agree with a simple model which takes into account the energy cost due to surface area at the edge of the pore. The energy cost at the edge of the pore is experimentally varied by using a lamellar-forming diblock copolymer membrane. The underlying molecular architecture causes increased frustration at the pore edge resulting in an enhanced cost of pore formation.

\end{abstract}

\maketitle
Nucleation and growth occurs in a variety of physical systems where there is a phase transition. Crystallization of ice, the formation of micelles in solution, and the growth of diamonds from vitreous carbon are all systems where a nucleus can form and then either grow or shrink depending on its size~\cite{Schmelzer2005}. In crystallization, for example, there is a trade-off between the volumetric free energy contribution which favors the crystalline phase, and the surface area dependent cost of the interfacial tension between the two phases which favors a single liquid phase~\cite{jones2002soft}.  By examining the free energy cost of creating a nucleated phase with radius $r$, the critical radius $r_c$ can be derived from classical nucleation theory. For a nucleus with $r<r_c$, the nucleated phase is unstable to fluctuations, while for $r>r_c$ the nucleated phase is energetically favorable and the nucleus can grow. The same physics governs diverse phenomena including bubble nucleation of false vacuum states in inflationary cosmology~\cite{Frampton1976}, as well as many biologically important processes such as amyloid-beta protein aggregation~\cite{Lomakin1996,Garai2008}, microtubule growth~\cite{Piehl2004}, and pore formation in cell membranes~\cite{phillips2012physical}. Membrane pores can be created by pathogenic bacteria to invade target cells~\cite{Yamashita2014}, and play an important role in many biological processes such as mechanical force transduction~\cite{Coste2012} and water permeability of biological membranes~\cite{Agre2002}. In this letter we focus on pore formation in a membrane.

For the nucleation of a pore in a membrane, there is a balance between two competing effects. First, there is an energy cost of having an interface between the membrane and its surrounding phase. Because the presence of a pore removes interface, this reduces the interfacial energy cost. Opposing this effect is the energetic cost of creating an edge around the perimeter of the pore. The stability and growth of pores depends on the relative contribution of interface reduction compared to the edge cost. The free energy cost of creating a pore in a membrane which has two surfaces with interfacial tension $\gamma$ is given by~\cite{Taupin1975}
\begin{equation}\label{eq:DeltaG}
\Delta G(r) =  2 \pi r \Gamma - 2 \pi r^2 \gamma,
\end{equation}
where $\Gamma$ is the edge tension (or line tension) of the pore: the free energy cost \emph{per unit length} of creating interface at the edge of the pore [see schematic in Fig.~1(a)]. The edge tension in lipid membranes has been the subject of both experimental~\cite{Zhelev1993,Sandre1999,Bier2002,Loi2002,Puech2003,Karatekin2003,Lee2004,Garcia-Saez2007,Lee2008} and	 theoretical~\cite{Netz1996,Muller1996,Talanquer2003,Leontiadou2004,Tolpekina2004,Wang2005,Li2013} studies, which address important questions about the composition profile of the lipid molecules around the pore edge~\cite{Muller1996,Wang2005}  as well as the role of peptides in the free energy of the pore~\cite{Lee2004,Garcia-Saez2007,Lee2008}. Studying pore formation in lipid membranes can be an experimental challenge due to their small size and fast timescales. 

\begin{figure}[tb]
	\includegraphics[width=1.0\columnwidth]{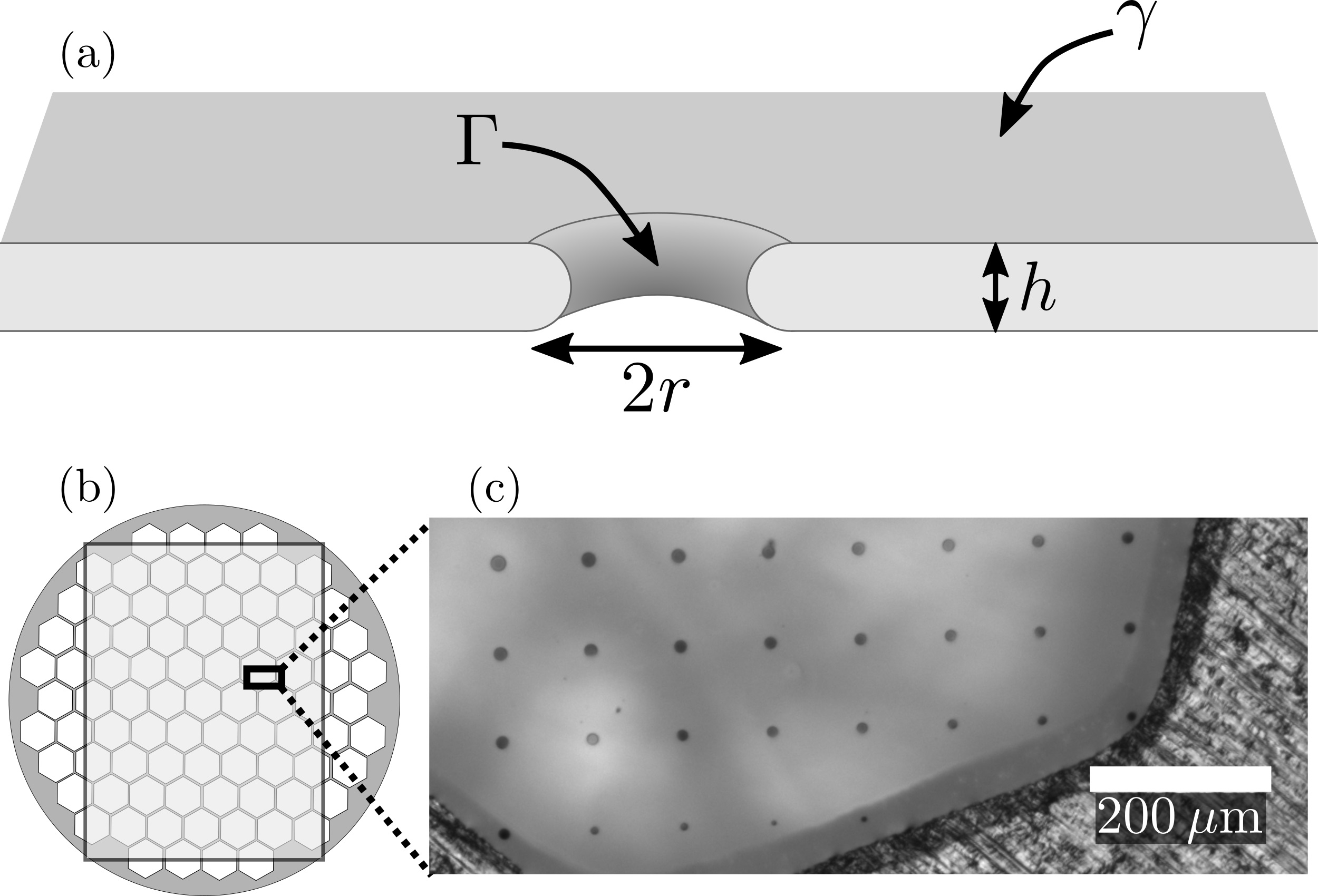}
	\caption{(a) Cross-sectional diagram of a pore with radius $r$ in a membrane with thickness $h$, interfacial tension $\gamma$, and edge tension $\Gamma$. (b) Schematic of a freestanding polymer membrane (light grey rectangle) supported by a stainless steel grid. Each hexagon in the grid is $\sim 1\,\mathrm{mm}$ across. (c) Optical microscopy image of a small portion of a membrane showing 29 pores with varying radius.\label{Fig1}}
\end{figure}

Here we study the formation of pores in a model membrane system, a thin freely-suspended polymer film. We use membranes which are hundreds of nanometers thick, allowing for direct measurement using optical or atomic force microscopy. Furthermore, because the viscosity of a polymer film depends strongly on temperature, the dynamics of pore formation can be tuned to convenient timescales. Previously, viscous polymer membranes have been used as a model system to study the growth dynamics of rupture in bubbles~\cite{Debregeas1998} as well as to study the spontaneous nucleation and growth of holes~\cite{Dalnoki-Veress1999a,Croll2010,Rathfon2011}. Our approach is to purposefully make holes of various sizes in a liquid polymer membrane and allow them to evolve under the influence of surface tension. Holes with a radius below a critical radius are dominated by the edge tension and shrink, while those larger than the critical radius grow. Examining the shrinking or growth of many holes allows for the determination of $r_c$ which depends on the film thickness $h$. Thus, we are able to watch the growth and shrinkage of pores, which gives a direct measurement of the critical pore size. Unlike previous studies which rely on indirect inference from fluid flow through the membrane and changing the mechanical tension of the membrane~\cite{Zhelev1993,Sandre1999}, our measurements provide a straightforward approach to probing the physics of equation (\ref{eq:DeltaG}) - a standard textbook equation - for which direct measurement has been elusive.

To create the polymer membranes, polystyrene (obtained from Scientific Polymer Products, USA, with number averaged molecular weight $M_n=47.5\,\mathrm{kg/mol}$, and polydispersity index 1.01) was spincast from dilute toluene solution onto freshly cleaved mica (Ted Pella, USA). The polystyrene (PS) films were then floated off the mica onto a deionized water bath (18.2 M$\Omega$ cm, Pall, USA), and transferred onto silicon (University Wafer, USA). Pores in the polystyrene film were created by a focused laser spike annealing setup~\cite{Singer2011,Singer2013,Benzaquen2015}, where a tightly focused laser (Coherent, Verdi V2, 532 nm) was used to locally heat the silicon wafer supporting the PS film. This heating had two effects. First, the local temperature of the PS film became greater than its glass transition temperature ($T_g \approx 100^{\circ{}}\mathrm{C}$), which created a local region of liquid PS. Second, the laser caused a lateral temperature gradient within the PS film with a corresponding surface tension gradient. The surface tension of PS decreases with increasing temperature, thus thermocapillary flow drives the formation of a hole in the PS film around the center of the laser beam. By varying the power and exposure time of the laser, holes of various sizes in the PS film could be created. The process of exposing the film to the laser, closing the laser shutter and then translating the sample to a new location was repeated to yield hundreds of holes in each PS film. Having hundreds of holes ensured a good statistical average for the critical radius measurement. After preparing the holes, the films were floated off the silicon back onto the water bath, and picked up onto stainless steel grids shown schematically in Fig.~1(b). The grid supports the PS film, with many freely-suspended regions of PS film. The films were then heated above $T_g$ ($T \gtrsim 110^{\circ{}}\mathrm{C}$) on a heating stage (Linkham, UK) to bring them into the liquid state. The size of the pores was measured \emph{in-situ} using optical microscopy (OM) [Olympus, USA] as shown in Fig.~1(c), or by intermittently quenching the samples to room temperature, deep into the glassy state of PS where no flow occurs, and scanning the film with atomic force microscopy (AFM) [Veeco Cailber, USA]~\cite{note}.

The AFM profiles of three pores in a film with thickness $h=810\,\mathrm{nm}$ are shown in Fig.~2 for three different annealing times $t$ at temperatures above $T_g$. The largest hole grew (left column). Remarkably, the the next largest was stable over the timescale of the experiment (middle column), and the smallest hole shrank (right column). The size of the stable pore gives an estimate of $r_c \approx 500\,\mathrm{nm}$ for $h=810\,\mathrm{nm}$. 
\begin{figure}[tb]
	\includegraphics[width=1.0\columnwidth]{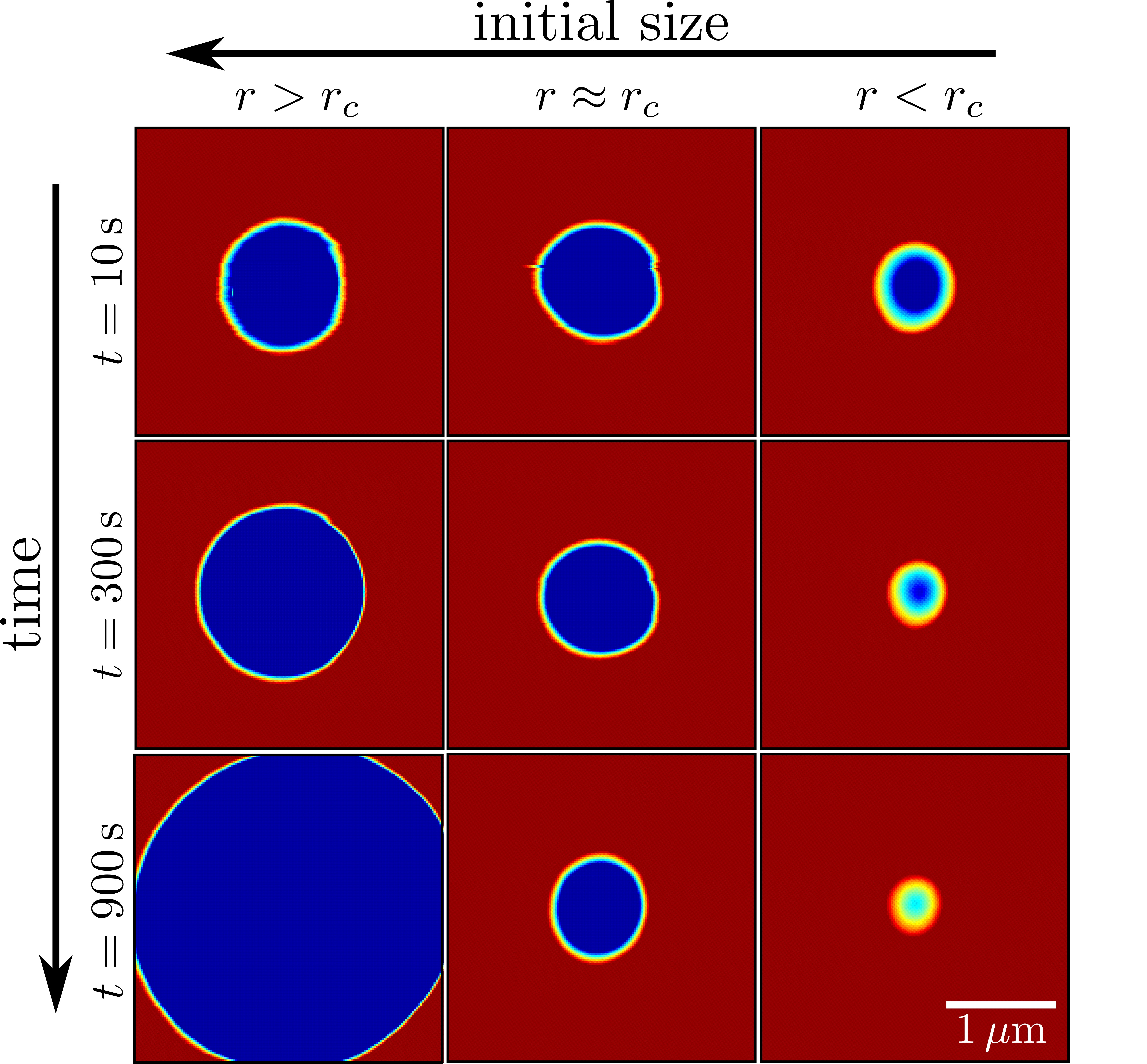}
	\caption{The time evolution of pores in a viscous polystyrene membrane ($h=810\,\mathrm{nm}$) as measured by AFM. The largest hole grows (left column), while the smallest hole closes (right column). An intermediate hole (middle column) is stable in size over the measurement. \label{Fig2}}
\end{figure}
In order to perform measurements on hundreds of pores in the same film, OM was used to measure the pore radius. Fig.~3 shows the time evolution of the radius of three different holes for a film $h\approx 900\,\mathrm{nm}$ thick. The largest pore grows exponentially with time, consistent with previous studies~\cite{Debregeas1995,Dalnoki-Veress1999a,Croll2010}. The intermediate sized pore does not change, while the smallest hole rapidly shrinks. Repeating this measurement on a sample with hundreds of holes, we find only 8 stable holes for the film with $h\approx 900\,\mathrm{nm}$. From the size of the stable holes, we find $r_c = 730\pm 100 \,\mathrm{nm}$ for $h=965\,\mathrm{nm}$.
\begin{figure}[tb]
	\includegraphics[width=1.0\columnwidth]{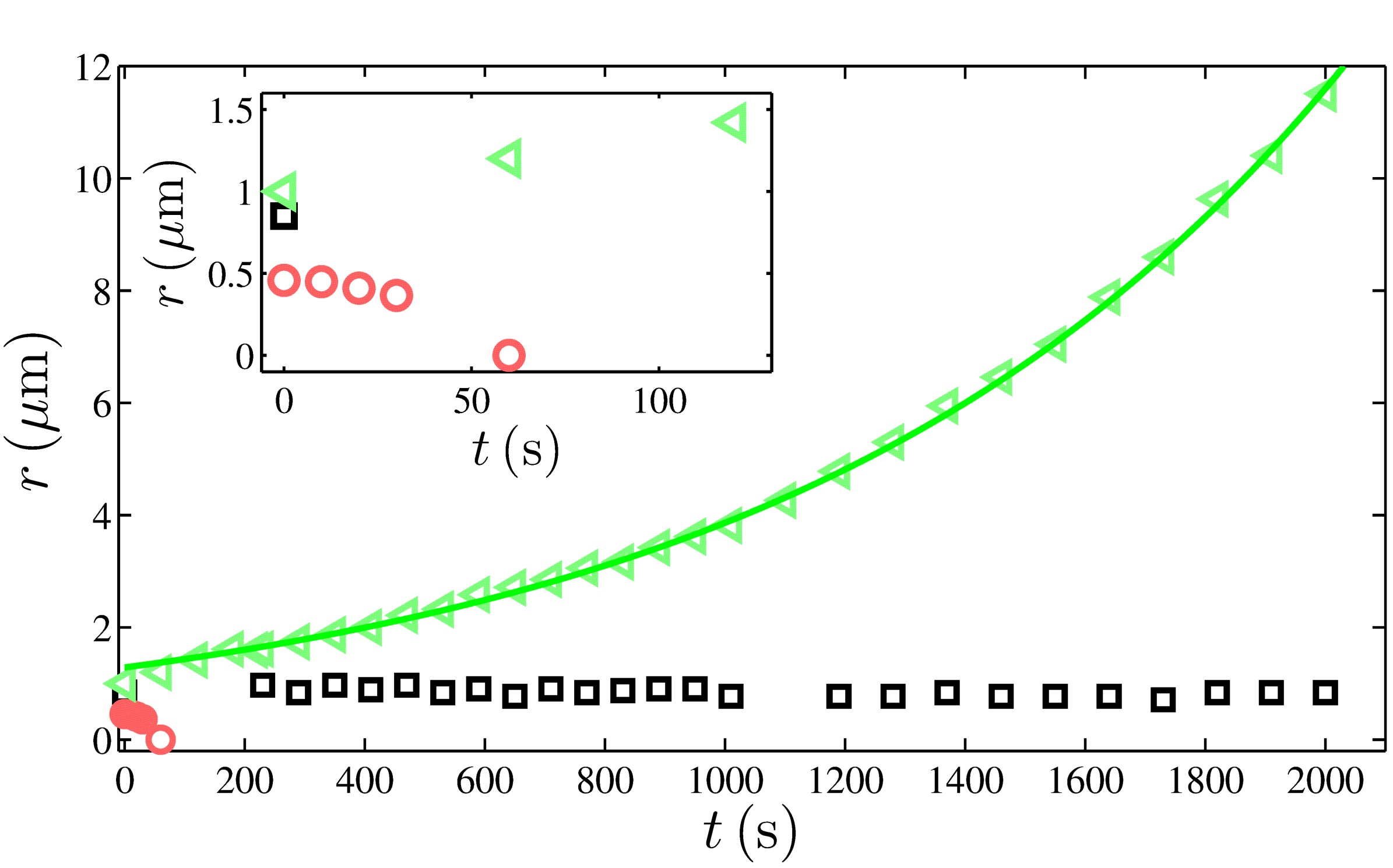}
	\caption{The radius of three different sized pores in a film $h\approx 900\,\mathrm{nm}$ thick. The pore which started off with the largest radius (triangles) grew over time, while the smallest hole (circles) shrank and closed. The intermediate sized pore (squares) was stable in size during the measurement. The solid line is a fit of $r=r_0 e^{t/\tau}$ to the growing hole, with $r_0 = 1.3\,\mathrm{\mu m}$ and $\tau = 910\,\mathrm{s}$. The inset shows the same data over a smaller range of time to highlight the dynamics of the shrinking pore.\label{Fig3}}
\end{figure}
The results for $9$ different PS film thicknesses show $r_c$ depends linearly on $h$ (Fig.~4).

\begin{figure}[b]
	\includegraphics[width=1.0\columnwidth]{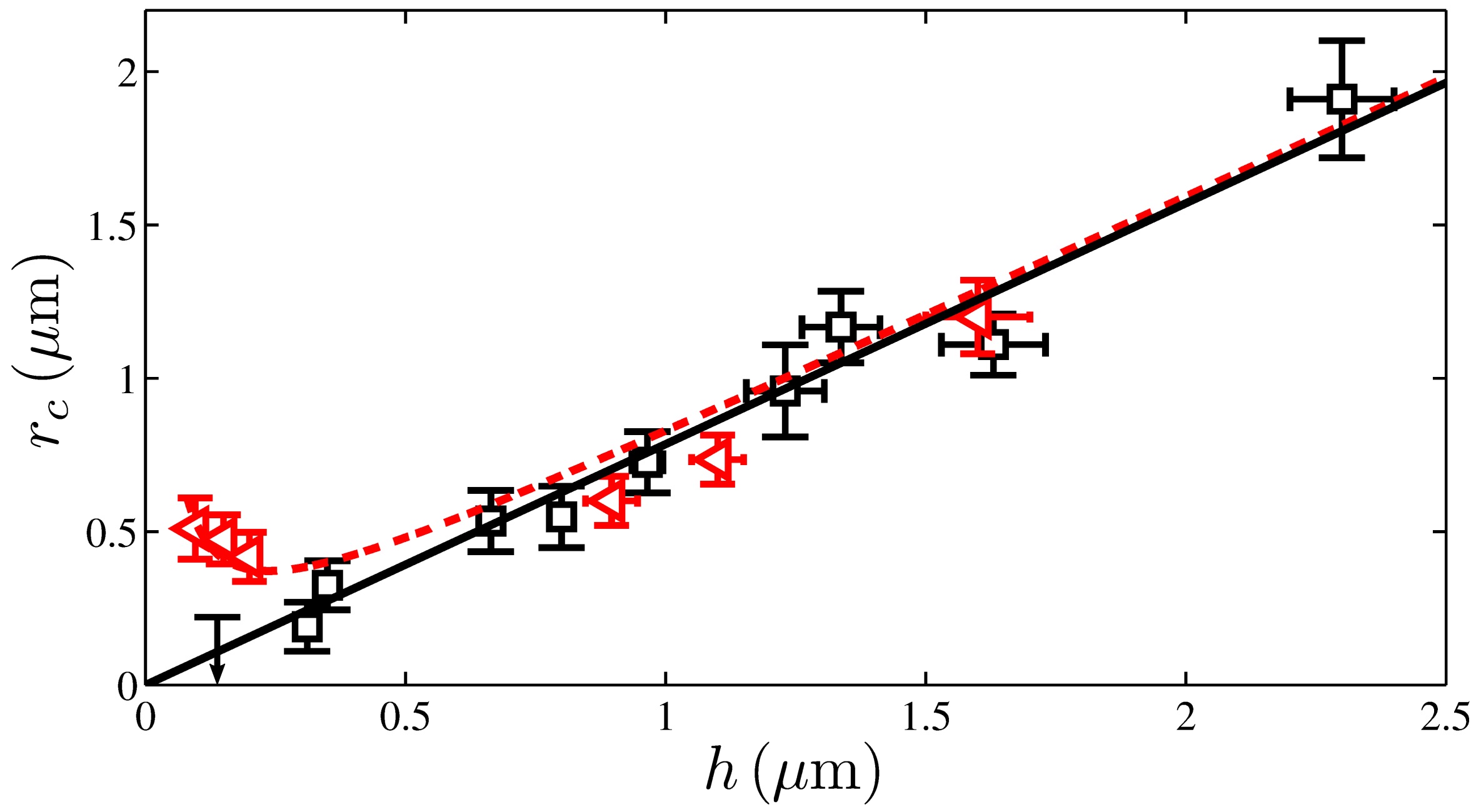}
	\caption{Film thickness dependence of the critical radius for a PS homopolymer (squares) and a PS-PMMA diblock copolymer with additional edge tension (triangles). The solid line is the theoretical prediction of Eq.~(\ref{eq:rc_homopolymer}) with no fitting parameters. The dashed line is a fit of Eq.~(\ref{eq:rc_diblock}) to the diblock copolymer data. For the thinnest homopolymer film used, all the measured pores grew, providing an upper bound for the critical radius at this film thickness, denoted by the downward arrow. \label{Fig4}}
\end{figure}

To understand the linear relation between $r_c$ and $h$, we need to consider the edge tension term in Eq.(\ref{eq:DeltaG}). Since the PS films are composed of a single polymeric species and the film thickness is much larger than typical molecular dimensions, the energy cost of creating the rim of the pore is just the bulk surface tension cost. If we consider the edge of the pore to be a torus as has been done in previous theoretical work~\cite{Wohlert2006,Li2013,safran2003statistical}, with a diameter $h$ [see schematic in Fig.~1(a)], then the surface area of the rim is given by the surface area of the inner half of a torus $A = \pi^2 h r - \pi h^2$~\cite{note2}. The free energy of the PS pore is then
\begin{equation} \label{eq:DeltaGhomopolymer}
\frac{\Delta G(r)}{\gamma h^2}= -\pi + \pi^2 \left(\frac{r}{h}\right)- 2 \pi \left(\frac{r}{h}\right)^2,
\end{equation}
where we identify the edge tension as
\begin{equation}\label{eq:homopolymerLT}
\Gamma_{\textrm{homopolymer}} = \gamma h\left(\frac{\pi}{2}  - \frac{h}{2r}\right).
\end{equation}The solid black line in Fig.~5 shows the free energy cost of a pore as given by Eq.~(\ref{eq:DeltaGhomopolymer}). The single maximum of the free energy is an energy barrier to the growth of a pore. This prevents smaller pores from growing, and ultimately leads to their closure. Pores which are left of the peak in the free energy shrink, while those to the right of the peak grow. The critical radius is given by the maximum in the free energy, which we find by setting $\partial_r \Delta G|_{r=r_c} = 0$, 
\begin{equation}\label{eq:rc_homopolymer}
r_c = \pi h/4.
\end{equation}

\begin{figure}[b]
	\includegraphics[width=0.9\columnwidth]{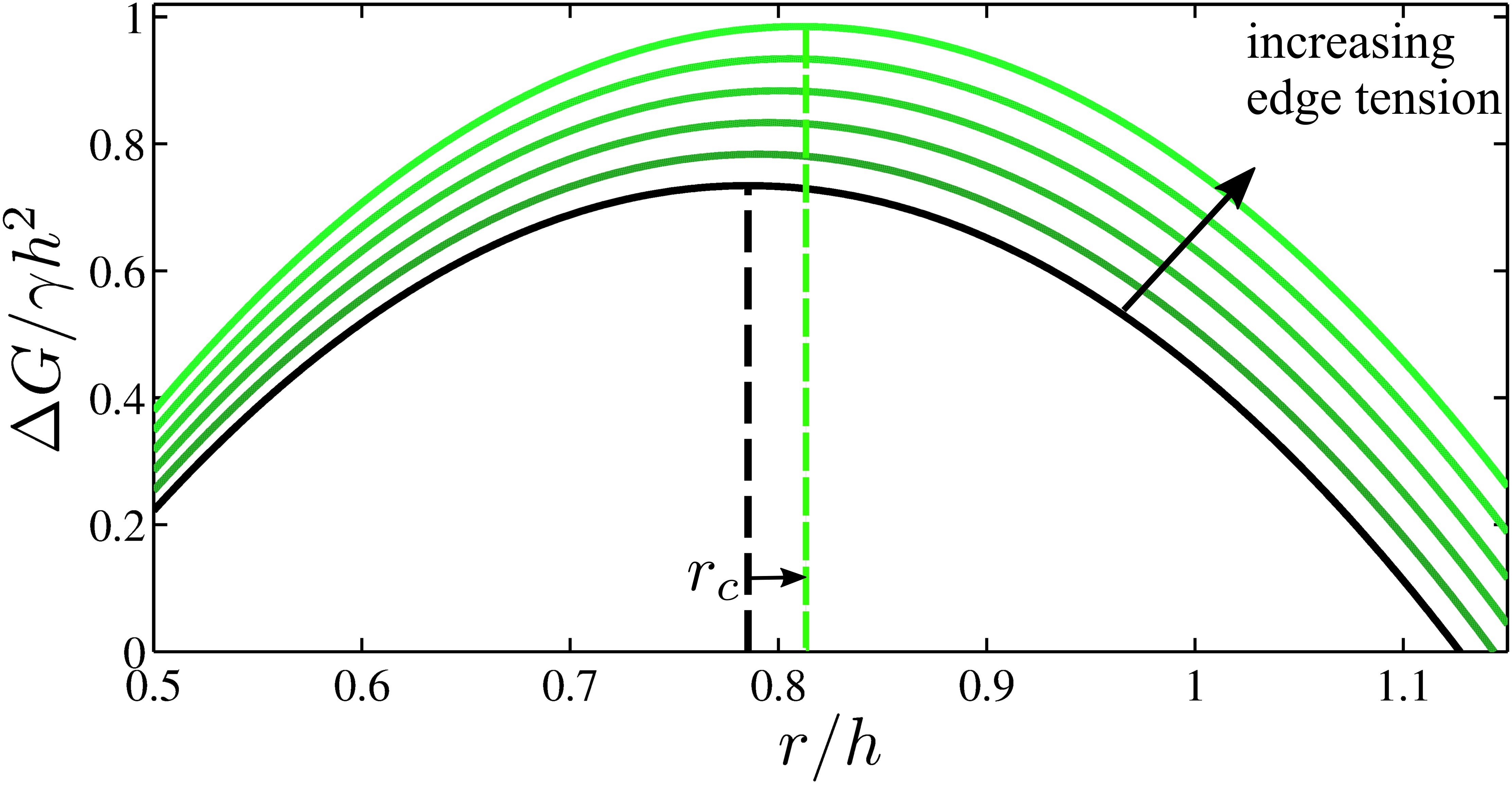}
	\caption{(color online). The normalized free energy cost of creating a pore with radius $r$ in a viscous membrane of film thickness $h$ and surface tension $\gamma$. The solid black curve is given by Eq.~(\ref{eq:DeltaG}), and the value of $r_c$ (dashed black line) is given by Eq.~(\ref{eq:rc_homopolymer}) for a film with no confinement effects. The green curves are the free energy cost of creating a pore when there is an additional contribution to the edge tension from the molecular structure.\label{Fig5}}
\end{figure}

The simple theoretical expression for the critical radius, Eq.(\ref{eq:rc_homopolymer}), is shown in Fig.~4 (solid line), and agrees with the experimentally measured values with no fitting parameters. Since $h>200\,\mathrm{nm}$ for all measured films, the contribution from disjoining pressure is negligible~\cite{Jacobs2008}. The excellent agreement between theory and experiment confirms the assumed geometry of the pore and the assumption that surface tension is the only free energy contribution in creating a pore in PS homopolymer films. The PS films are idealized model membranes with only the total surface area contributing to the free energy.

Now that we have a well-characterized model membrane, we turn to incorporating one complicating aspect associated with biological membranes. The stability of pores in membranes is largely determined by the effects of edge tension. For example, peptide binding can alter the edge tension which stabilizes pores~\cite{Garcia-Saez2007}. In contrast, here we explore the effect of altering the edge tension in a simplified manner. Instead of using a homopolymer, we use a diblock copolymer for the polymer membrane. A diblock copolymer is a type of polymer in which each molecule contains two distinct chemical blocks~\cite{jones2002soft}. Due to an enthalpic repulsion of the two parts of each chain, phase separation is favourable at sufficiently low temperatures. But because the two chemical species are covalently bonded together, phase separation can only occur on the lengthscale of the individual molecules (typically $10-100\,\mathrm{nm}$). Here we use a symmetric polystyrene-poly(methyl methacrylate) (PS-b-PMMA) diblock copolymer (with $M_n = 25\,\mathrm{kg/mol}$ for the PS block and $M_n = 26\,\mathrm{kg/mol}$ for the PMMA block, obtained from Polymer Source, Canada), and follow the same sample preparation and experimental procedure used for the homopolymer. A symmetric diblock copolymer will phase separate into lamellar layers with equilibrium thickness $L$, similar to a lipid bilayer. In fact, diblock copolymers have been used to model lipid bilayer membranes in previous theoretical work~\cite{Li2013}. When confined to a thin film, there is an extra free energy cost of a curved diblock surface~\cite{Wilmes2006}. Unlike the case of a homopolymer membrane where the free energy cost at the edge of the pore is just due to the bulk surface tension of the fluid, a diblock copolymer has an added energy cost of rearranging molecules near the curved edge of the pore which disrupts the flat lamellar geometry. We assume that this cost is inversely related the radius of curvature. Here, the radius of curvature is $h/2$, and to first order this adds an additional cost to the edge tension of a hole given by Eq.~(\ref{eq:homopolymerLT}) in terms of the non-dimensional curvature $L/h$ as 
\begin{equation} \label{eq:diblockLT}
\Gamma_{\textrm{diblock}} = \Gamma_{\textrm{homopolymer}}+ \phi \frac{L}{h},
\end{equation} 
where $\phi$ is an energy cost per unit length associated with the curvature. Combining Eq.~(\ref{eq:diblockLT}) and Eq.~(\ref{eq:DeltaG}) gives
\begin{equation} \label{eq:DeltaGdiblock}
\frac{\Delta G(r)}{\gamma h^2}= - \pi + \left(\pi^2 + \frac{2 \pi \phi L}{\gamma h^2} \right)\left(\frac{r}{h}\right)  - 2 \pi \left(\frac{r}{h}\right)^2.
\end{equation} 

The free energy cost of a pore in a system with an added edge tension contribution is shown in Fig.~5. The free energy given by Eq.~(\ref{eq:DeltaGdiblock}) is plotted for successively larger values of the edge tension. The larger edge tension results in an increased free energy barrier to hole growth, which shifts $r_c$ to progressively higher values. 
To obtain the critical radius for the diblock copolymer case, we once again set $\partial_r \Delta G|_{r=r_c} = 0$, and obtain the critical radius
\begin{equation}\label{eq:rc_diblock}
r_c = \pi h/4+ \frac{\phi L }{2 \gamma h}.
\end{equation}
To test Eq.~(\ref{eq:rc_diblock}), diblock copolymer membranes with pores were created in the same way as the PS homopolymer case, and the critical radius was measured for varying film thicknesses (triangles in Fig.~4). For thick films ($>800\,\mathrm{nm}$), the critical radius measured in the diblock films show no significant difference from that of the homopolymer films. In thinner films, the effect of the additional edge tension of the diblock copolymer film becomes important, and the critical radius begins to increase as the film thickness is decreased. The dashed line is the fit of Eq.~(\ref{eq:rc_diblock}) to the diblock data with $\phi L /\gamma = 0.09 \pm 0.04 \,\mathrm{\mu m^2}$. This ratio defines a characteristic lengthscale $h^\star = \sqrt{\phi L/\gamma} = 300 \pm 70\,\mathrm{nm}$. The contribution to the edge tension from the structure of the diblock copolymer becomes dominant when the second term in Eq.~(\ref{eq:rc_diblock}) is larger than the first term, which occurs when $h \lesssim h^\star$. The observed deviation of the diblock copolymer from the homopolymer case in Fig.~4, is consistent with the molecular structure being significant for films thinner than $h^\star$.

Polymer membranes have provided a convenient system to examine a nucleated process. The ability to use direct optical measurements allows for a conceptually straightforward demonstration of nucleation and growth, which are important in many other physical processes. The simplicity of the system allows for an uncomplicated mathematical description of the physics involved. The ability to tune the pore stability by changing molecular architecture gives a clear understanding of the competing contributions to the free energy of the system. 

In conclusion, we have studied the stability of pores in a model membrane. We find that for a simple polystyrene film, the critical radius depends linearly on film thickness. A model which only accounts for the surface tension contribution to the free energy and takes into consideration the surface area at the edge of the pore describes our experimental results with no free parameters. On the other hand, we find that for a membrane made of a diblock copolymer the critical radius does not match the simple model for thin films. The ordered microstructure of the diblock causes deviations from the bulk free energy cost due to surface tension. This added contribution to the edge tension can be accommodated with an additional term in the free energy which becomes important for thin films and is able to describe the experimental diblock copolymer critical radius with one fitting parameter. The study of pores in polymer membranes provides a model system for nucleation and growth where the critical radius can be derived from simple mathematical arguments. In this model membrane system the effect of the edge tension, which is important to biological processes, can be altered simply by changing the molecular architecture of the polymer used in the membrane.

\begin{acknowledgments}
Financial Support for this work was provided in part by NSERC (Canada). The authors would like to thank John Dutcher for valuable input.
\end{acknowledgments}


\begin{thebibliography}{41}%
	\makeatletter
	\providecommand \@ifxundefined [1]{%
		\@ifx{#1\undefined}
	}%
	\providecommand \@ifnum [1]{%
		\ifnum #1\expandafter \@firstoftwo
		\else \expandafter \@secondoftwo
		\fi
	}%
	\providecommand \@ifx [1]{%
		\ifx #1\expandafter \@firstoftwo
		\else \expandafter \@secondoftwo
		\fi
	}%
	\providecommand \natexlab [1]{#1}%
	\providecommand \enquote  [1]{``#1''}%
	\providecommand \bibnamefont  [1]{#1}%
	\providecommand \bibfnamefont [1]{#1}%
	\providecommand \citenamefont [1]{#1}%
	\providecommand \href@noop [0]{\@secondoftwo}%
	\providecommand \href [0]{\begingroup \@sanitize@url \@href}%
	\providecommand \@href[1]{\@@startlink{#1}\@@href}%
	\providecommand \@@href[1]{\endgroup#1\@@endlink}%
	\providecommand \@sanitize@url [0]{\catcode `\\12\catcode `\$12\catcode
		`\&12\catcode `\#12\catcode `\^12\catcode `\_12\catcode `\%12\relax}%
	\providecommand \@@startlink[1]{}%
	\providecommand \@@endlink[0]{}%
	\providecommand \url  [0]{\begingroup\@sanitize@url \@url }%
	\providecommand \@url [1]{\endgroup\@href {#1}{\urlprefix }}%
	\providecommand \urlprefix  [0]{URL }%
	\providecommand \Eprint [0]{\href }%
	\providecommand \doibase [0]{http://dx.doi.org/}%
	\providecommand \selectlanguage [0]{\@gobble}%
	\providecommand \bibinfo  [0]{\@secondoftwo}%
	\providecommand \bibfield  [0]{\@secondoftwo}%
	\providecommand \translation [1]{[#1]}%
	\providecommand \BibitemOpen [0]{}%
	\providecommand \bibitemStop [0]{}%
	\providecommand \bibitemNoStop [0]{.\EOS\space}%
	\providecommand \EOS [0]{\spacefactor3000\relax}%
	\providecommand \BibitemShut  [1]{\csname bibitem#1\endcsname}%
	\let\auto@bib@innerbib\@empty
	%</preamble>
	\bibitem [{\citenamefont {Schmelzer}\ \emph {et~al.}(2005)\citenamefont
		{Schmelzer}, \citenamefont {R{\"{o}}pke},\ and\ \citenamefont
		{Priezzhev}}]{Schmelzer2005}%
	\BibitemOpen
	\bibfield  {author} {\bibinfo {author} {\bibfnamefont {J.}~\bibnamefont
			{Schmelzer}}, \bibinfo {author} {\bibfnamefont {G.}~\bibnamefont
			{R{\"{o}}pke}}, \ and\ \bibinfo {author} {\bibfnamefont {V.}~\bibnamefont
			{Priezzhev}},\ }\href
	{http://onlinelibrary.wiley.com/doi/10.1002/3527604790.fmatter/summary}
	{\emph {\bibinfo {title} {{Nucleation theory and applications}}}}\ (\bibinfo
	{year} {2005})\BibitemShut {NoStop}%
	\bibitem [{\citenamefont {Jones}(2002)}]{jones2002soft}%
	\BibitemOpen
	\bibfield  {author} {\bibinfo {author} {\bibfnamefont {R.~A.~L.}\
			\bibnamefont {Jones}},\ }\href@noop {} {\emph {\bibinfo {title} {{Soft
					condensed matter}}}}\ (\bibinfo  {publisher} {Oxford University Press},\
	\bibinfo {address} {New York},\ \bibinfo {year} {2002})\BibitemShut {NoStop}%
	\bibitem [{\citenamefont {Frampton}(1976)}]{Frampton1976}%
	\BibitemOpen
	\bibfield  {author} {\bibinfo {author} {\bibfnamefont {P.~H.}\ \bibnamefont
			{Frampton}},\ }\href {\doibase 10.1103/PhysRevLett.37.1378} {\bibfield
		{journal} {\bibinfo  {journal} {Phys. Rev. Lett.}\ }\textbf {\bibinfo
			{volume} {37}},\ \bibinfo {pages} {1378} (\bibinfo {year}
		{1976})}\BibitemShut {NoStop}%
	\bibitem [{\citenamefont {Lomakin}\ \emph {et~al.}(1996)\citenamefont
		{Lomakin}, \citenamefont {Chung}, \citenamefont {Benedek}, \citenamefont
		{Kirschner},\ and\ \citenamefont {Teplow}}]{Lomakin1996}%
	\BibitemOpen
	\bibfield  {author} {\bibinfo {author} {\bibfnamefont {A.}~\bibnamefont
			{Lomakin}}, \bibinfo {author} {\bibfnamefont {D.~S.}\ \bibnamefont {Chung}},
		\bibinfo {author} {\bibfnamefont {G.~B.}\ \bibnamefont {Benedek}}, \bibinfo
		{author} {\bibfnamefont {D.~A.}\ \bibnamefont {Kirschner}}, \ and\ \bibinfo
		{author} {\bibfnamefont {D.~B.}\ \bibnamefont {Teplow}},\ }\href {\doibase
		10.1073/pnas.93.3.1125} {\bibfield  {journal} {\bibinfo  {journal} {Proc.
				Natl. Acad. Sci. U. S. A.}\ }\textbf {\bibinfo {volume} {93}},\ \bibinfo
		{pages} {1125} (\bibinfo {year} {1996})}\BibitemShut {NoStop}%
	\bibitem [{\citenamefont {Garai}\ \emph {et~al.}(2008)\citenamefont {Garai},
		\citenamefont {Sahoo}, \citenamefont {Sengupta},\ and\ \citenamefont
		{Maiti}}]{Garai2008}%
	\BibitemOpen
	\bibfield  {author} {\bibinfo {author} {\bibfnamefont {K.}~\bibnamefont
			{Garai}}, \bibinfo {author} {\bibfnamefont {B.}~\bibnamefont {Sahoo}},
		\bibinfo {author} {\bibfnamefont {P.}~\bibnamefont {Sengupta}}, \ and\
		\bibinfo {author} {\bibfnamefont {S.}~\bibnamefont {Maiti}},\ }\href
	{\doibase 10.1063/1.2822322} {\bibfield  {journal} {\bibinfo  {journal} {J.
				Chem. Phys.}\ }\textbf {\bibinfo {volume} {128}} (\bibinfo {year} {2008}),\
		10.1063/1.2822322}\BibitemShut {NoStop}%
	\bibitem [{\citenamefont {Piehl}\ \emph {et~al.}(2004)\citenamefont {Piehl},
		\citenamefont {Tulu}, \citenamefont {Wadsworth},\ and\ \citenamefont
		{Cassimeris}}]{Piehl2004}%
	\BibitemOpen
	\bibfield  {author} {\bibinfo {author} {\bibfnamefont {M.}~\bibnamefont
			{Piehl}}, \bibinfo {author} {\bibfnamefont {U.~S.}\ \bibnamefont {Tulu}},
		\bibinfo {author} {\bibfnamefont {P.}~\bibnamefont {Wadsworth}}, \ and\
		\bibinfo {author} {\bibfnamefont {L.}~\bibnamefont {Cassimeris}},\ }\href
	{\doibase 10.1073/pnas.0308205100} {\bibfield  {journal} {\bibinfo  {journal}
			{Proc. Natl. Acad. Sci. U. S. A.}\ }\textbf {\bibinfo {volume} {101}},\
		\bibinfo {pages} {1584} (\bibinfo {year} {2004})}\BibitemShut {NoStop}%
	\bibitem [{\citenamefont {Phillips}\ \emph {et~al.}(2012)\citenamefont
		{Phillips}, \citenamefont {Kondev}, \citenamefont {Theriot},\ and\
		\citenamefont {Garcia}}]{phillips2012physical}%
	\BibitemOpen
	\bibfield  {author} {\bibinfo {author} {\bibfnamefont {R.}~\bibnamefont
			{Phillips}}, \bibinfo {author} {\bibfnamefont {J.}~\bibnamefont {Kondev}},
		\bibinfo {author} {\bibfnamefont {J.}~\bibnamefont {Theriot}}, \ and\
		\bibinfo {author} {\bibfnamefont {H.}~\bibnamefont {Garcia}},\ }\href@noop {}
	{\emph {\bibinfo {title} {{Physical biology of the cell}}}}\ (\bibinfo
	{publisher} {Garland Science},\ \bibinfo {year} {2012})\BibitemShut {NoStop}%
	\bibitem [{\citenamefont {Yamashita}\ \emph {et~al.}(2014)\citenamefont
		{Yamashita}, \citenamefont {Sugawara}, \citenamefont {Takeshita},
		\citenamefont {Kaneko}, \citenamefont {Kamio}, \citenamefont {Tanaka},
		\citenamefont {Tanaka},\ and\ \citenamefont {Yao}}]{Yamashita2014}%
	\BibitemOpen
	\bibfield  {author} {\bibinfo {author} {\bibfnamefont {D.}~\bibnamefont
			{Yamashita}}, \bibinfo {author} {\bibfnamefont {T.}~\bibnamefont {Sugawara}},
		\bibinfo {author} {\bibfnamefont {M.}~\bibnamefont {Takeshita}}, \bibinfo
		{author} {\bibfnamefont {J.}~\bibnamefont {Kaneko}}, \bibinfo {author}
		{\bibfnamefont {Y.}~\bibnamefont {Kamio}}, \bibinfo {author} {\bibfnamefont
			{I.}~\bibnamefont {Tanaka}}, \bibinfo {author} {\bibfnamefont
			{Y.}~\bibnamefont {Tanaka}}, \ and\ \bibinfo {author} {\bibfnamefont
			{M.}~\bibnamefont {Yao}},\ }\href {\doibase 10.1038/ncomms5897} {\bibfield
		{journal} {\bibinfo  {journal} {Nat. Commun.}\ }\textbf {\bibinfo {volume}
			{5}},\ \bibinfo {pages} {4897} (\bibinfo {year} {2014})}\BibitemShut
	{NoStop}%
	\bibitem [{\citenamefont {Coste}\ \emph {et~al.}(2012)\citenamefont {Coste},
		\citenamefont {Xiao}, \citenamefont {Santos}, \citenamefont {Syeda},
		\citenamefont {Grandl}, \citenamefont {Spencer}, \citenamefont {Kim},
		\citenamefont {Schmidt}, \citenamefont {Mathur}, \citenamefont {Dubin},
		\citenamefont {Montal},\ and\ \citenamefont {Patapoutian}}]{Coste2012}%
	\BibitemOpen
	\bibfield  {author} {\bibinfo {author} {\bibfnamefont {B.}~\bibnamefont
			{Coste}}, \bibinfo {author} {\bibfnamefont {B.}~\bibnamefont {Xiao}},
		\bibinfo {author} {\bibfnamefont {J.~S.}\ \bibnamefont {Santos}}, \bibinfo
		{author} {\bibfnamefont {R.}~\bibnamefont {Syeda}}, \bibinfo {author}
		{\bibfnamefont {J.}~\bibnamefont {Grandl}}, \bibinfo {author} {\bibfnamefont
			{K.~S.}\ \bibnamefont {Spencer}}, \bibinfo {author} {\bibfnamefont {S.~E.}\
			\bibnamefont {Kim}}, \bibinfo {author} {\bibfnamefont {M.}~\bibnamefont
			{Schmidt}}, \bibinfo {author} {\bibfnamefont {J.}~\bibnamefont {Mathur}},
		\bibinfo {author} {\bibfnamefont {A.~E.}\ \bibnamefont {Dubin}}, \bibinfo
		{author} {\bibfnamefont {M.}~\bibnamefont {Montal}}, \ and\ \bibinfo {author}
		{\bibfnamefont {A.}~\bibnamefont {Patapoutian}},\ }\href {\doibase
		10.1038/nature10812} {\bibfield  {journal} {\bibinfo  {journal} {Nature}\
		}\textbf {\bibinfo {volume} {483}},\ \bibinfo {pages} {176} (\bibinfo {year}
		{2012})}\BibitemShut {NoStop}%
	\bibitem [{\citenamefont {Agre}\ \emph {et~al.}(2002)\citenamefont {Agre},
		\citenamefont {King}, \citenamefont {Yasui}, \citenamefont {Guggino},
		\citenamefont {Ottersen}, \citenamefont {Fujiyoshi}, \citenamefont {Engel},\
		and\ \citenamefont {Nielsen}}]{Agre2002}%
	\BibitemOpen
	\bibfield  {author} {\bibinfo {author} {\bibfnamefont {P.}~\bibnamefont
			{Agre}}, \bibinfo {author} {\bibfnamefont {L.~S.}\ \bibnamefont {King}},
		\bibinfo {author} {\bibfnamefont {M.}~\bibnamefont {Yasui}}, \bibinfo
		{author} {\bibfnamefont {W.~B.}\ \bibnamefont {Guggino}}, \bibinfo {author}
		{\bibfnamefont {O.~P.}\ \bibnamefont {Ottersen}}, \bibinfo {author}
		{\bibfnamefont {Y.}~\bibnamefont {Fujiyoshi}}, \bibinfo {author}
		{\bibfnamefont {A.}~\bibnamefont {Engel}}, \ and\ \bibinfo {author}
		{\bibfnamefont {S.}~\bibnamefont {Nielsen}},\ }\href {\doibase
		10.1113/jphysiol.2002.020818} {\bibfield  {journal} {\bibinfo  {journal} {J.
				Physiol.}\ }\textbf {\bibinfo {volume} {542}},\ \bibinfo {pages} {3}
		(\bibinfo {year} {2002})}\BibitemShut {NoStop}%
	\bibitem [{\citenamefont {Taupin}\ \emph {et~al.}(1975)\citenamefont {Taupin},
		\citenamefont {Dvolaitzky},\ and\ \citenamefont {Sauterey}}]{Taupin1975}%
	\BibitemOpen
	\bibfield  {author} {\bibinfo {author} {\bibfnamefont {C.}~\bibnamefont
			{Taupin}}, \bibinfo {author} {\bibfnamefont {M.}~\bibnamefont {Dvolaitzky}},
		\ and\ \bibinfo {author} {\bibfnamefont {C.}~\bibnamefont {Sauterey}},\
	}\href@noop {} {\bibfield  {journal} {\bibinfo  {journal} {Biochemistry}\
	}\textbf {\bibinfo {volume} {14}},\ \bibinfo {pages} {4771} (\bibinfo {year}
	{1975})}\BibitemShut {NoStop}%
\bibitem [{\citenamefont {Zhelev}\ and\ \citenamefont
	{Needham}(1993)}]{Zhelev1993}%
\BibitemOpen
\bibfield  {author} {\bibinfo {author} {\bibfnamefont {D.~V.}\ \bibnamefont
		{Zhelev}}\ and\ \bibinfo {author} {\bibfnamefont {D.}~\bibnamefont
		{Needham}},\ }\href {\doibase 10.1016/0005-2736(93)90319-U} {\bibfield
	{journal} {\bibinfo  {journal} {Biochim. Biophys. Acta - Biomembr.}\ }\textbf
	{\bibinfo {volume} {1147}},\ \bibinfo {pages} {89} (\bibinfo {year}
	{1993})}\BibitemShut {NoStop}%
\bibitem [{\citenamefont {Sandre}\ \emph {et~al.}(1999)\citenamefont {Sandre},
	\citenamefont {Moreaux},\ and\ \citenamefont {Brochard-Wyart}}]{Sandre1999}%
\BibitemOpen
\bibfield  {author} {\bibinfo {author} {\bibfnamefont {O.}~\bibnamefont
		{Sandre}}, \bibinfo {author} {\bibfnamefont {L.}~\bibnamefont {Moreaux}}, \
	and\ \bibinfo {author} {\bibfnamefont {F.}~\bibnamefont {Brochard-Wyart}},\
}\href {\doibase 10.1073/pnas.96.19.10591} {\bibfield  {journal} {\bibinfo
	{journal} {Proc. Natl. Acad. Sci. U. S. A.}\ }\textbf {\bibinfo {volume}
	{96}},\ \bibinfo {pages} {10591} (\bibinfo {year} {1999})}\BibitemShut
{NoStop}%
\bibitem [{\citenamefont {Bier}\ \emph {et~al.}(2002)\citenamefont {Bier},
	\citenamefont {Chen}, \citenamefont {Gowrishankar}, \citenamefont
	{Astumian},\ and\ \citenamefont {Lee}}]{Bier2002}%
\BibitemOpen
\bibfield  {author} {\bibinfo {author} {\bibfnamefont {M.}~\bibnamefont
		{Bier}}, \bibinfo {author} {\bibfnamefont {W.}~\bibnamefont {Chen}}, \bibinfo
	{author} {\bibfnamefont {T.~R.}\ \bibnamefont {Gowrishankar}}, \bibinfo
	{author} {\bibfnamefont {R.~D.}\ \bibnamefont {Astumian}}, \ and\ \bibinfo
	{author} {\bibfnamefont {R.~C.}\ \bibnamefont {Lee}},\ }\href {\doibase
	10.1103/PhysRevE.66.062905} {\bibfield  {journal} {\bibinfo  {journal} {Phys.
			Rev. E}\ }\textbf {\bibinfo {volume} {66}},\ \bibinfo {pages} {062905}
	(\bibinfo {year} {2002})}\BibitemShut {NoStop}%
\bibitem [{\citenamefont {Loi}\ \emph {et~al.}(2002)\citenamefont {Loi},
	\citenamefont {Sun}, \citenamefont {Franz},\ and\ \citenamefont
	{Butt}}]{Loi2002}%
\BibitemOpen
\bibfield  {author} {\bibinfo {author} {\bibfnamefont {S.}~\bibnamefont
		{Loi}}, \bibinfo {author} {\bibfnamefont {G.}~\bibnamefont {Sun}}, \bibinfo
	{author} {\bibfnamefont {V.}~\bibnamefont {Franz}}, \ and\ \bibinfo {author}
	{\bibfnamefont {H.-J.}\ \bibnamefont {Butt}},\ }\href {\doibase
	10.1103/PhysRevE.66.031602} {\bibfield  {journal} {\bibinfo  {journal} {Phys.
			Rev. E}\ }\textbf {\bibinfo {volume} {66}},\ \bibinfo {pages} {031602}
	(\bibinfo {year} {2002})}\BibitemShut {NoStop}%
\bibitem [{\citenamefont {Puech}\ \emph {et~al.}(2003)\citenamefont {Puech},
	\citenamefont {Borghi}, \citenamefont {Karatekin},\ and\ \citenamefont
	{Brochard-Wyart}}]{Puech2003}%
\BibitemOpen
\bibfield  {author} {\bibinfo {author} {\bibfnamefont {P.-H.}\ \bibnamefont
		{Puech}}, \bibinfo {author} {\bibfnamefont {N.}~\bibnamefont {Borghi}},
	\bibinfo {author} {\bibfnamefont {E.}~\bibnamefont {Karatekin}}, \ and\
	\bibinfo {author} {\bibfnamefont {F.}~\bibnamefont {Brochard-Wyart}},\ }\href
{\doibase 10.1103/PhysRevLett.90.128304} {\bibfield  {journal} {\bibinfo
		{journal} {Phys. Rev. Lett.}\ }\textbf {\bibinfo {volume} {90}},\ \bibinfo
	{pages} {128304} (\bibinfo {year} {2003})}\BibitemShut {NoStop}%
\bibitem [{\citenamefont {Karatekin}\ \emph {et~al.}(2003)\citenamefont
	{Karatekin}, \citenamefont {Sandre}, \citenamefont {Guitouni}, \citenamefont
	{Borghi}, \citenamefont {Puech},\ and\ \citenamefont
	{Brochard-Wyart}}]{Karatekin2003}%
\BibitemOpen
\bibfield  {author} {\bibinfo {author} {\bibfnamefont {E.}~\bibnamefont
		{Karatekin}}, \bibinfo {author} {\bibfnamefont {O.}~\bibnamefont {Sandre}},
	\bibinfo {author} {\bibfnamefont {H.}~\bibnamefont {Guitouni}}, \bibinfo
	{author} {\bibfnamefont {N.}~\bibnamefont {Borghi}}, \bibinfo {author}
	{\bibfnamefont {P.-H.}\ \bibnamefont {Puech}}, \ and\ \bibinfo {author}
	{\bibfnamefont {F.}~\bibnamefont {Brochard-Wyart}},\ }\href {\doibase
	10.1016/S0006-3495(03)74981-9} {\bibfield  {journal} {\bibinfo  {journal}
		{Biophys. J.}\ }\textbf {\bibinfo {volume} {84}},\ \bibinfo {pages} {1734}
	(\bibinfo {year} {2003})}\BibitemShut {NoStop}%
\bibitem [{\citenamefont {Lee}\ \emph {et~al.}(2004)\citenamefont {Lee},
	\citenamefont {Chen},\ and\ \citenamefont {Huang}}]{Lee2004}%
\BibitemOpen
\bibfield  {author} {\bibinfo {author} {\bibfnamefont {M.~T.}\ \bibnamefont
		{Lee}}, \bibinfo {author} {\bibfnamefont {F.~Y.}\ \bibnamefont {Chen}}, \
	and\ \bibinfo {author} {\bibfnamefont {H.~W.}\ \bibnamefont {Huang}},\ }\href
{\doibase 10.1021/bi036153r} {\bibfield  {journal} {\bibinfo  {journal}
		{Biochemistry}\ }\textbf {\bibinfo {volume} {43}},\ \bibinfo {pages} {3590}
	(\bibinfo {year} {2004})}\BibitemShut {NoStop}%
\bibitem [{\citenamefont {Garc{\'{\i}}a-S{\'{a}}ez}\ \emph
	{et~al.}(2007)\citenamefont {Garc{\'{\i}}a-S{\'{a}}ez}, \citenamefont
	{Chiantia}, \citenamefont {Salgado},\ and\ \citenamefont
	{Schwille}}]{Garcia-Saez2007}%
\BibitemOpen
\bibfield  {author} {\bibinfo {author} {\bibfnamefont {A.~J.}\ \bibnamefont
		{Garc{\'{\i}}a-S{\'{a}}ez}}, \bibinfo {author} {\bibfnamefont
		{S.}~\bibnamefont {Chiantia}}, \bibinfo {author} {\bibfnamefont
		{J.}~\bibnamefont {Salgado}}, \ and\ \bibinfo {author} {\bibfnamefont
		{P.}~\bibnamefont {Schwille}},\ }\href {\doibase 10.1529/biophysj.106.100370}
{\bibfield  {journal} {\bibinfo  {journal} {Biophys. J.}\ }\textbf {\bibinfo
		{volume} {93}},\ \bibinfo {pages} {103} (\bibinfo {year} {2007})}\BibitemShut
{NoStop}%
\bibitem [{\citenamefont {Lee}\ \emph {et~al.}(2008)\citenamefont {Lee},
	\citenamefont {Hung}, \citenamefont {Chen},\ and\ \citenamefont
	{Huang}}]{Lee2008}%
\BibitemOpen
\bibfield  {author} {\bibinfo {author} {\bibfnamefont {M.-T.}\ \bibnamefont
		{Lee}}, \bibinfo {author} {\bibfnamefont {W.-C.}\ \bibnamefont {Hung}},
	\bibinfo {author} {\bibfnamefont {F.-Y.}\ \bibnamefont {Chen}}, \ and\
	\bibinfo {author} {\bibfnamefont {H.~W.}\ \bibnamefont {Huang}},\ }\href@noop
{} {\bibfield  {journal} {\bibinfo  {journal} {Proc. Natl. Acad. Sci. U. S.
			A.}\ }\textbf {\bibinfo {volume} {105}},\ \bibinfo {pages} {5087} (\bibinfo
	{year} {2008})}\BibitemShut {NoStop}%
\bibitem [{\citenamefont {Netz}\ and\ \citenamefont {Schick}(1996)}]{Netz1996}%
\BibitemOpen
\bibfield  {author} {\bibinfo {author} {\bibfnamefont {R.}~\bibnamefont
		{Netz}}\ and\ \bibinfo {author} {\bibfnamefont {M.}~\bibnamefont {Schick}},\
}\href {\doibase 10.1103/PhysRevE.53.3875} {\bibfield  {journal} {\bibinfo
	{journal} {Phys. Rev. E}\ }\textbf {\bibinfo {volume} {53}},\ \bibinfo
{pages} {3875} (\bibinfo {year} {1996})}\BibitemShut {NoStop}%
\bibitem [{\citenamefont {MuÌˆller}\ and\ \citenamefont
	{Schick}(1996)}]{Muller1996}%
\BibitemOpen
\bibfield  {author} {\bibinfo {author} {\bibfnamefont {M.}~\bibnamefont
		{MuÌˆller}}\ and\ \bibinfo {author} {\bibfnamefont {M.}~\bibnamefont
		{Schick}},\ }\href {\doibase 10.1063/1.472682} {\bibfield  {journal}
	{\bibinfo  {journal} {J. Chem. Phys.}\ }\textbf {\bibinfo {volume} {105}},\
	\bibinfo {pages} {8282} (\bibinfo {year} {1996})}\BibitemShut {NoStop}%
\bibitem [{\citenamefont {Talanquer}\ and\ \citenamefont
	{Oxtoby}(2003)}]{Talanquer2003}%
\BibitemOpen
\bibfield  {author} {\bibinfo {author} {\bibfnamefont {V.}~\bibnamefont
		{Talanquer}}\ and\ \bibinfo {author} {\bibfnamefont {D.~W.}\ \bibnamefont
		{Oxtoby}},\ }\href {\doibase 10.1063/1.1526093} {\bibfield  {journal}
	{\bibinfo  {journal} {J. Chem. Phys.}\ }\textbf {\bibinfo {volume} {118}},\
	\bibinfo {pages} {872} (\bibinfo {year} {2003})}\BibitemShut {NoStop}%
\bibitem [{\citenamefont {Leontiadou}\ \emph {et~al.}(2004)\citenamefont
	{Leontiadou}, \citenamefont {Mark},\ and\ \citenamefont
	{Marrink}}]{Leontiadou2004}%
\BibitemOpen
\bibfield  {author} {\bibinfo {author} {\bibfnamefont {H.}~\bibnamefont
		{Leontiadou}}, \bibinfo {author} {\bibfnamefont {A.~E.}\ \bibnamefont
		{Mark}}, \ and\ \bibinfo {author} {\bibfnamefont {S.~J.}\ \bibnamefont
		{Marrink}},\ }\href {\doibase 10.1016/S0006-3495(04)74275-7} {\bibfield
	{journal} {\bibinfo  {journal} {Biophys. J.}\ }\textbf {\bibinfo {volume}
		{86}},\ \bibinfo {pages} {2156} (\bibinfo {year} {2004})}\BibitemShut
{NoStop}%
\bibitem [{\citenamefont {Tolpekina}\ \emph {et~al.}(2004)\citenamefont
	{Tolpekina}, \citenamefont {den Otter},\ and\ \citenamefont
	{Briels}}]{Tolpekina2004}%
\BibitemOpen
\bibfield  {author} {\bibinfo {author} {\bibfnamefont {T.~V.}\ \bibnamefont
		{Tolpekina}}, \bibinfo {author} {\bibfnamefont {W.~K.}\ \bibnamefont {den
			Otter}}, \ and\ \bibinfo {author} {\bibfnamefont {W.~J.}\ \bibnamefont
		{Briels}},\ }\href {\doibase 10.1063/1.1815296} {\bibfield  {journal}
	{\bibinfo  {journal} {J. Chem. Phys.}\ }\textbf {\bibinfo {volume} {121}},\
	\bibinfo {pages} {12060} (\bibinfo {year} {2004})}\BibitemShut {NoStop}%
\bibitem [{\citenamefont {Wang}\ and\ \citenamefont
	{Frenkel}(2005)}]{Wang2005}%
\BibitemOpen
\bibfield  {author} {\bibinfo {author} {\bibfnamefont {Z.-J.}\ \bibnamefont
		{Wang}}\ and\ \bibinfo {author} {\bibfnamefont {D.}~\bibnamefont {Frenkel}},\
}\href {\doibase 10.1063/1.2060666} {\bibfield  {journal} {\bibinfo
	{journal} {J. Chem. Phys.}\ }\textbf {\bibinfo {volume} {123}},\ \bibinfo
{pages} {154701} (\bibinfo {year} {2005})}\BibitemShut {NoStop}%
\bibitem [{\citenamefont {Li}\ \emph {et~al.}(2013)\citenamefont {Li},
	\citenamefont {Pastor}, \citenamefont {Shi}, \citenamefont {Schmid},\ and\
	\citenamefont {Zhou}}]{Li2013}%
\BibitemOpen
\bibfield  {author} {\bibinfo {author} {\bibfnamefont {J.}~\bibnamefont
		{Li}}, \bibinfo {author} {\bibfnamefont {K.~A.}\ \bibnamefont {Pastor}},
	\bibinfo {author} {\bibfnamefont {A.~C.}\ \bibnamefont {Shi}}, \bibinfo
	{author} {\bibfnamefont {F.}~\bibnamefont {Schmid}}, \ and\ \bibinfo {author}
	{\bibfnamefont {J.}~\bibnamefont {Zhou}},\ }\href {\doibase
	10.1103/PhysRevE.88.012718} {\bibfield  {journal} {\bibinfo  {journal} {Phys.
			Rev. E - Stat. Nonlinear, Soft Matter Phys.}\ }\textbf {\bibinfo {volume}
		{88}},\ \bibinfo {pages} {1} (\bibinfo {year} {2013})}\BibitemShut {NoStop}%
\bibitem [{\citenamefont {Debregeas}\ \emph {et~al.}(1998)\citenamefont
	{Debregeas}, \citenamefont {de~Gennes},\ and\ \citenamefont
	{Brochard-Wyart}}]{Debregeas1998}%
\BibitemOpen
\bibfield  {author} {\bibinfo {author} {\bibfnamefont {G.}~\bibnamefont
		{Debregeas}}, \bibinfo {author} {\bibfnamefont {P.}~\bibnamefont
		{de~Gennes}}, \ and\ \bibinfo {author} {\bibfnamefont {F.}~\bibnamefont
		{Brochard-Wyart}},\ }\href {\doibase 10.1126/science.279.5357.1704}
{\bibfield  {journal} {\bibinfo  {journal} {Science (80-. ).}\ }\textbf
	{\bibinfo {volume} {279}},\ \bibinfo {pages} {1704} (\bibinfo {year}
	{1998})}\BibitemShut {NoStop}%
\bibitem [{Note1()}]{note}%
\BibitemOpen
\bibinfo {note} {Using AFM to measure pores in a freely-suspended membrane is technically challenging. Vibrations can be easily induced in the membrane, which we remedied by acoustically insulating the AFM. As well, when the AFM tip scans over a pore it loses contact with the membrane. As a result, the  vertical piezo of the AFM extends to its maximum range, pushing the AFM tip  through the pore by several microns. This large extension can cause the edge  of the AFM tip to crash into the edge of the pore, causing damage to the  membrane. To avoid damaging the membrane, we operated the AFM close to its  maximum vertical extension when in contact with the membrane. Thus, when the AFM tip was rastered over a pore, the tip only extended a small distance into  the pore.}\BibitemShut {Stop}%
\bibitem [{\citenamefont {Dalnoki-Veress}\ \emph {et~al.}(1999)\citenamefont
	{Dalnoki-Veress}, \citenamefont {Nickel}, \citenamefont {Roth},\ and\
	\citenamefont {Dutcher}}]{Dalnoki-Veress1999a}%
\BibitemOpen
\bibfield  {author} {\bibinfo {author} {\bibfnamefont {K.}~\bibnamefont
		{Dalnoki-Veress}}, \bibinfo {author} {\bibfnamefont {B.}~\bibnamefont
		{Nickel}}, \bibinfo {author} {\bibfnamefont {C.}~\bibnamefont {Roth}}, \ and\
	\bibinfo {author} {\bibfnamefont {J.}~\bibnamefont {Dutcher}},\ }\href
{\doibase 10.1103/PhysRevE.59.2153} {\bibfield  {journal} {\bibinfo
		{journal} {Phys. Rev. E}\ }\textbf {\bibinfo {volume} {59}},\ \bibinfo
	{pages} {2153} (\bibinfo {year} {1999})}\BibitemShut {NoStop}%
\bibitem [{\citenamefont {Croll}\ and\ \citenamefont
	{Dalnoki-Veress}(2010)}]{Croll2010}%
\BibitemOpen
\bibfield  {author} {\bibinfo {author} {\bibfnamefont {A.~B.}\ \bibnamefont
		{Croll}}\ and\ \bibinfo {author} {\bibfnamefont {K.}~\bibnamefont
		{Dalnoki-Veress}},\ }\href {\doibase 10.1039/c0sm00253d} {\bibfield
	{journal} {\bibinfo  {journal} {Soft Matter}\ }\textbf {\bibinfo {volume}
		{6}},\ \bibinfo {pages} {5547} (\bibinfo {year} {2010})}\BibitemShut
{NoStop}%
\bibitem [{\citenamefont {Rathfon}\ \emph {et~al.}(2011)\citenamefont
	{Rathfon}, \citenamefont {Cohn}, \citenamefont {Crosby},\ and\ \citenamefont
	{Tew}}]{Rathfon2011}%
\BibitemOpen
\bibfield  {author} {\bibinfo {author} {\bibfnamefont {J.~M.}\ \bibnamefont
		{Rathfon}}, \bibinfo {author} {\bibfnamefont {R.~W.}\ \bibnamefont {Cohn}},
	\bibinfo {author} {\bibfnamefont {A.~J.}\ \bibnamefont {Crosby}}, \ and\
	\bibinfo {author} {\bibfnamefont {G.~N.}\ \bibnamefont {Tew}},\ }\href
{\doibase 10.1021/ma1020227} {\bibfield  {journal} {\bibinfo  {journal}
		{Macromolecules}\ }\textbf {\bibinfo {volume} {44}},\ \bibinfo {pages} {134}
	(\bibinfo {year} {2011})}\BibitemShut {NoStop}%
\bibitem [{\citenamefont {Singer}\ \emph {et~al.}(2011)\citenamefont {Singer},
	\citenamefont {Kooi},\ and\ \citenamefont {Thomas}}]{Singer2011}%
\BibitemOpen
\bibfield  {author} {\bibinfo {author} {\bibfnamefont {J.~P.}\ \bibnamefont
		{Singer}}, \bibinfo {author} {\bibfnamefont {S.~E.}\ \bibnamefont {Kooi}}, \
	and\ \bibinfo {author} {\bibfnamefont {E.~L.}\ \bibnamefont {Thomas}},\
}\href {\doibase 10.1039/c1nr10050e} {\bibfield  {journal} {\bibinfo
	{journal} {Nanoscale}\ }\textbf {\bibinfo {volume} {3}},\ \bibinfo {pages}
{2730} (\bibinfo {year} {2011})}\BibitemShut {NoStop}%
\bibitem [{\citenamefont {Singer}\ \emph {et~al.}(2013)\citenamefont {Singer},
	\citenamefont {Lin}, \citenamefont {Kooi}, \citenamefont {Kimerling},
	\citenamefont {Michel},\ and\ \citenamefont {Thomas}}]{Singer2013}%
\BibitemOpen
\bibfield  {author} {\bibinfo {author} {\bibfnamefont {J.~P.}\ \bibnamefont
		{Singer}}, \bibinfo {author} {\bibfnamefont {P.-T.}\ \bibnamefont {Lin}},
	\bibinfo {author} {\bibfnamefont {S.~E.}\ \bibnamefont {Kooi}}, \bibinfo
	{author} {\bibfnamefont {L.~C.}\ \bibnamefont {Kimerling}}, \bibinfo {author}
	{\bibfnamefont {J.}~\bibnamefont {Michel}}, \ and\ \bibinfo {author}
	{\bibfnamefont {E.~L.}\ \bibnamefont {Thomas}},\ }\href {\doibase
	10.1002/adma.201302777} {\bibfield  {journal} {\bibinfo  {journal} {Adv.
			Mater.}\ }\textbf {\bibinfo {volume} {25}},\ \bibinfo {pages} {6100}
	(\bibinfo {year} {2013})}\BibitemShut {NoStop}%
\bibitem [{\citenamefont {Benzaquen}\ \emph {et~al.}(2015)\citenamefont
	{Benzaquen}, \citenamefont {Ilton}, \citenamefont {Massa}, \citenamefont
	{Salez}, \citenamefont {Fowler}, \citenamefont {Rapha{\"{e}}l},\ and\
	\citenamefont {Dalnoki-Veress}}]{Benzaquen2015}%
\BibitemOpen
\bibfield  {author} {\bibinfo {author} {\bibfnamefont {M.}~\bibnamefont
		{Benzaquen}}, \bibinfo {author} {\bibfnamefont {M.}~\bibnamefont {Ilton}},
	\bibinfo {author} {\bibfnamefont {M.~V.}\ \bibnamefont {Massa}}, \bibinfo
	{author} {\bibfnamefont {T.}~\bibnamefont {Salez}}, \bibinfo {author}
	{\bibfnamefont {P.}~\bibnamefont {Fowler}}, \bibinfo {author} {\bibfnamefont
		{E.}~\bibnamefont {Rapha{\"{e}}l}}, \ and\ \bibinfo {author} {\bibfnamefont
		{K.}~\bibnamefont {Dalnoki-Veress}},\ }\href {\doibase 10.1063/1.4927599}
{\bibfield  {journal} {\bibinfo  {journal} {Appl. Phys. Lett.}\ }\textbf
	{\bibinfo {volume} {107}},\ \bibinfo {pages} {053103} (\bibinfo {year}
	{2015})}\BibitemShut {NoStop}%
\bibitem [{\citenamefont {Debr{\'{e}}geas}\ \emph {et~al.}(1995)\citenamefont
	{Debr{\'{e}}geas}, \citenamefont {Martin},\ and\ \citenamefont
	{Brochard-Wyart}}]{Debregeas1995}%
\BibitemOpen
\bibfield  {author} {\bibinfo {author} {\bibfnamefont {G.}~\bibnamefont
		{Debr{\'{e}}geas}}, \bibinfo {author} {\bibfnamefont {P.}~\bibnamefont
		{Martin}}, \ and\ \bibinfo {author} {\bibfnamefont {F.}~\bibnamefont
		{Brochard-Wyart}},\ }\href {\doibase 10.1103/PhysRevLett.75.3886} {\bibfield
	{journal} {\bibinfo  {journal} {Phys. Rev. Lett.}\ }\textbf {\bibinfo
		{volume} {75}},\ \bibinfo {pages} {3886} (\bibinfo {year}
	{1995})}\BibitemShut {NoStop}%
\bibitem [{\citenamefont {Wohlert}\ \emph {et~al.}(2006)\citenamefont
	{Wohlert}, \citenamefont {{Den Otter}}, \citenamefont {Edholm},\ and\
	\citenamefont {Briels}}]{Wohlert2006}%
\BibitemOpen
\bibfield  {author} {\bibinfo {author} {\bibfnamefont {J.}~\bibnamefont
		{Wohlert}}, \bibinfo {author} {\bibfnamefont {W.~K.}\ \bibnamefont {{Den
				Otter}}}, \bibinfo {author} {\bibfnamefont {O.}~\bibnamefont {Edholm}}, \
	and\ \bibinfo {author} {\bibfnamefont {W.~J.}\ \bibnamefont {Briels}},\
}\href {\doibase 10.1063/1.2171965} {\bibfield  {journal} {\bibinfo
	{journal} {J. Chem. Phys.}\ }\textbf {\bibinfo {volume} {124}} (\bibinfo
{year} {2006}),\ 10.1063/1.2171965}\BibitemShut {NoStop}%
\bibitem [{\citenamefont {Safran}(2003)}]{safran2003statistical}%
\BibitemOpen
\bibfield  {author} {\bibinfo {author} {\bibfnamefont {S.~A.}\ \bibnamefont
		{Safran}},\ }\href@noop {} {\emph {\bibinfo {title} {{Statistical
				Thermodynamics of Surfaces, Interfaces, and Membranes}}}},\ Frontiers in
physics\ (\bibinfo  {publisher} {Westview Press},\ \bibinfo {year}
{2003})\BibitemShut {NoStop}%
\bibitem [{Note2()}]{note2}%
\BibitemOpen
\bibinfo {note} {The assumption that the edge of the pore is toroidal in shape is consistent with AFM measurements and no significant rim can be observed at the edge of the hole.}\BibitemShut {Stop}%

\bibitem [{\citenamefont {Jacobs}\ \emph {et~al.}(2008)\citenamefont {Jacobs},
	\citenamefont {Seemann},\ and\ \citenamefont {Herminghaus}}]{Jacobs2008}%
\BibitemOpen
\bibfield  {author} {\bibinfo {author} {\bibfnamefont {K.}~\bibnamefont
		{Jacobs}}, \bibinfo {author} {\bibfnamefont {R.}~\bibnamefont {Seemann}}, \
	and\ \bibinfo {author} {\bibfnamefont {S.}~\bibnamefont {Herminghaus}},\ }in\
\href@noop {} {\emph {\bibinfo {booktitle} {Polym. Thin Film.}}},\ \bibinfo
{editor} {edited by\ \bibinfo {editor} {\bibfnamefont {O.~K.~C.}\
		\bibnamefont {Tsui}}\ and\ \bibinfo {editor} {\bibfnamefont {T.~P.}\
		\bibnamefont {Russell}}}\ (\bibinfo  {publisher} {World Scientific},\
\bibinfo {year} {2008})\ Chap.~\bibinfo {chapter} {10}, pp.\ \bibinfo {pages}
{243--265}\BibitemShut {NoStop}%
\bibitem [{\citenamefont {Wilmes}\ \emph {et~al.}(2006)\citenamefont {Wilmes},
	\citenamefont {Durkee}, \citenamefont {Balsara},\ and\ \citenamefont
	{Liddle}}]{Wilmes2006}%
\BibitemOpen
\bibfield  {author} {\bibinfo {author} {\bibfnamefont {G.~M.}\ \bibnamefont
		{Wilmes}}, \bibinfo {author} {\bibfnamefont {D.~a.}\ \bibnamefont {Durkee}},
	\bibinfo {author} {\bibfnamefont {N.~P.}\ \bibnamefont {Balsara}}, \ and\
	\bibinfo {author} {\bibfnamefont {J.~A.}\ \bibnamefont {Liddle}},\ }\href
{\doibase 10.1021/ma0526443} {\bibfield  {journal} {\bibinfo  {journal}
		{Macromolecules}\ }\textbf {\bibinfo {volume} {39}},\ \bibinfo {pages} {2435}
	(\bibinfo {year} {2006})}\BibitemShut {NoStop}%
\bibitem [{\citenamefont {Croll}\ \emph {et~al.}(2006)\citenamefont {Croll},
	\citenamefont {Massa}, \citenamefont {Matsen},\ and\ \citenamefont
	{Dalnoki-Veress}}]{Croll2006}%
\BibitemOpen
\bibfield  {author} {\bibinfo {author} {\bibfnamefont {A.}~\bibnamefont
		{Croll}}, \bibinfo {author} {\bibfnamefont {M.}~\bibnamefont {Massa}},
	\bibinfo {author} {\bibfnamefont {M.}~\bibnamefont {Matsen}}, \ and\ \bibinfo
	{author} {\bibfnamefont {K.}~\bibnamefont {Dalnoki-Veress}},\ }\href
{\doibase 10.1103/PhysRevLett.97.204502} {\bibfield  {journal} {\bibinfo
		{journal} {Phys. Rev. Lett.}\ }\textbf {\bibinfo {volume} {97}},\ \bibinfo
	{pages} {1} (\bibinfo {year} {2006})}\BibitemShut {NoStop}%
\end{thebibliography}
\end{document}